\documentclass{llncs}
\usepackage{xspace}
\usepackage{latexml}
\usepackage{wrapfig}
\usepackage{listings}
\usepackage{amsmath}
\usepackage{url}
\lstdefinestyle{xml}{basicstyle=\scriptsize\sffamily,language=xml,escapechar=@,frame=tb}%
\newcommand{\defint}[4]{\int_{#1}^{#2} #3 d{#4}}
\newcommand{\xml}[1]{{\small\sffamily #1}}
\newcommand{\attr}[1]{{\small\sffamily #1}}
\newcommand{\xmath}{{\small\sffamily XMath}}
\title{Strategies for Parallel Markup}
\author{Bruce~R.~Miller\inst{1}
\thanks{The final publication is available at http://link.springer.com}}
\institute{National Institute of Standards and Technology, Gaithersburg, MD, USA}
\date{\today}

\begin{document}

\maketitle
\begin{abstract} 
Cross-referenced parallel markup for mathematics allows the combination of both
presentation and content representations while associating the components of each.
Interesting applications are enabled by such an arrangement, such as interaction
with parts of the presentation to manipulate and querying the corresponding content,
and enhanced search indexing.
Although the idea of such markup is hardly new, effective techniques for
creating and manipulating it are more difficult than it appears.
Since the structures and tokens in the two formats often do not correspond one-to-one,
decisions and heuristics must be developed to determine in which way each component
refers to and is referred to by components of the other representation.
Conversion between fine and coarse grained parallel markup complicates ID assignments.
In this paper, we will describe the techniques developed for \LaTeXML,
a \TeX/\LaTeX to XML converter, to create cross-referenced parallel MathML.
While we do not yet consider \LaTeXML's content MathML to be useful,
the current effort is a step towards that continuing goal.
\end{abstract}

%Keywords: MathML, parallel markup, \TeX\ conversion

\section{Introduction}
Parallel markup for mathematics provides the capability of providing
alternative representations of the mathematical expression, in particular,
both the presentation form of the mathematics, i.e. its appearance,
along with the content form, i.e. it's meaning or semantics.
Cross-linking between the two forms provides the connection between
them such that one can determine the meaning associated with
every visible fragment of the presentation and, conversely,
the visible manifestation of each semantic sub-expression. 
Thus cross-linked parallel markup provides not only the benefits of 
of the presentation and content forms, individually, but support
many other applications such as: hybrid search where both the presentation
and content can be taken into account simultaneously; interactive
applications where the visual representation forms part of the user-interface,
but supports computations based on the content representation.

Of course, the \emph{idea} of parallel markup is hardly new.
The \xml{m:semantics} element has been part of the MathML specification\cite{CarlisleEd:MathML3:base}
since the first version, in 1998!
What seems to be missing are effective strategies for creating,
manipulating and using this markup.
Fine-grained parallelism is when the smallest sub-expressions are represented
in multiple forms, whereas with coarse-grained parallelism
the entire expression appears in the several forms.
Fine-grained parallelism is generally easier to create initially,
and particularly when one deals with complex `transfix' notations,
or wants to preserve the appearance, but can infer the semantic intent, of each sub-expression.
Coarse-grained is often required by applications which may understand only a single format,
or are unable to disentangle the fined-grained structure.
HTML5 only just barely accepts coarse-grained parallel markup, for example.
Conversion from fine to coarse grained is not inherently difficult,
it can be carried out by a suitable walk of the expression tree
for each format. But what isn't so clear is how to maintain the associations
between the symbols (or more generally, the nodes) in the two trees.
Indeed, since there is typically no one-to-one correspondence between
the elements of each format. Fine-grained parallelism, by itself, doesn't guarantee a clear
association between all the symbols between the branches.

Our context here is \LaTeXML, a converter from \TeX/\LaTeX\ to
\XML, and thence to web appropriate formats such as \HTML, \MathML\ and OpenMath.
Input documents range from highly semantic markup such as sTeX\cite{Kohlhase:ulsmf08},
to intermediate such as used in DLMF\cite{MilYou:tadlmf02},
 to fairly undisciplined, purely presentational,
markup as found on arXiv\cite{StaKoh:tlcspx10}.
\TeX\ induces high expectations for quality formatting forcing us to preserve
the presentation of math. Meanwhile, the promise of global digital mathematics
libraries and the potential reuse of a legacy of mathematics material encourages
us to push as far as possible the extraction of content from such documents.
At the very least, we should preserve whatever semantics is available
in order to enable other technologies and research, such as LLaMaPuN\cite{GinJucAnc:alsaacl09},
to resolve the remaining ambiguities.

In this paper, we describe the markup used in \LaTeXML\ both for
macros with known semantics, and for the result of parsing,
and strategies for conversion to cross-linked, parallel markup
combining Presentation MathML (pMML) and Content MathML (cMML).
It should be noted that this does not mean that \LaTeXML\ is producing
useful quality cMML; the current work is a stepping stone towards
that long-term goal.

\section{Motivation}\label{motivation}
%\begin{wrapfigure}{l}{3in}
\begin{lstlisting}[style=xml,label=lst:basic.xml,
  float=t,caption={Internal representation of $a+F(a,b)$, after parsing (assuming $F$ as a function)}]
<XMApp>
  <XMTok meaning="plus" role="ADDOP">+</XMTok>
  <XMTok role="ID" font="italic">a</XMTok>
  <XMDual>
    <XMApp>
      <XMRef idref="m1.1"/>
      <XMRef idref="m1.2"/>
      <XMRef idref="m1.3"/>
    </XMApp>
    <XMApp>
      <XMTok role="FUNCTION" xml:id="m1.1" font="italic">F</XMTok>
      <XMWrap>
        <XMTok role="OPEN" stretchy="false">(</XMTok>
        <XMTok role="ID" xml:id="m1.2" font="italic">a</XMTok>
        <XMTok role="PUNCT">,</XMTok>
        <XMTok role="ID" xml:id="m1.3" font="italic">b</XMTok>
        <XMTok role="CLOSE" stretchy="false">)</XMTok>
      </XMWrap>
    </XMApp>
  </XMDual>
</XMApp>
\end{lstlisting}
Before diving into examples, a brief introduction to \LaTeXML's internal mathematics markup,
informally called \xmath, is in order.
This markup, inspired by OpenMath and both pMML and cMML,
is intentionally hybrid in order to capture both the
presentation and content properties of the mathematical objects
throughout the step-wise processing from raw \TeX\ markup, through parsing and,
ultimately, semantic annotation.
Please see the online manual\footnote{\url{http://dlmf.nist.gov/LaTeXML/manual/}}
for more details.
\begin{description}
\item[\xml{XMApp}] generalized application (think \xml{m:apply} or \xml{om:OMA});
\item[\xml{XMTok}] generalized token (think \xml{m:mi}, \xml{m:mo}, \xml{m:mn}, \xml{m:csymbol});
\item[\xml{XMDual}] parallel markup container of the content and presentation branches;
\item[\xml{XMRef}] shares nodes between branches of \xml{XMDual},
via \attr{xml:id} and \attr{idref} attributes;
\item[\xml{XMWrap}] container unparsed sequences of tokens or subtrees (think \xml{m:mrow}).
\end{description}

By way of motivation, consider the simple example in Listing \ref{lst:basic.xml}.
The \attr{role} attribute on tokens indicates the syntactic role that it plays in the grammar;
in this case, we've asserted that $F$ is a function, allowing
the expression to be parsed.  At the top-level, the sum requires no special parallel
treatment since the presentation for infix operators is trivially derived from
the content form (i.e. the application of `+' to its arguments).
The application of $F$ to its arguments benefits somewhat from parallel markup.
This is a typical situation with the fine-grained \xml{XMDual}:
the content branch is the application
of some function or operator (here $F$) to arguments (here $a$, $b$),
but they are represented indirectly using \xml{XMRef} to point to the corresponding
sub-expressions within the presentation.
While one could represent the delimiters and punctuation as attributes
(as in MathML's \xml{m:mfenced}), that loses attributes of those attributes
such as stretchiness, size or even color. A more compelling case is made
when more complex transfix notations or semantic macros are involved, as we will shortly see.

\begin{lstlisting}[style=xml,label=lst:basic.mathml,
  float=t,caption={\MathML\ representation of $a+F(a,b)$}]
<math display="block" alttext="a+F(a,b)" class="ltx_Math" id="m1">
  <semantics id="m1a">
    <mrow xref="m1.7.cmml" id="m1.7">
      <mi xref="m1.4.cmml" id="m1.4">a</mi>
      <mo xref="m1.5.cmml" id="m1.5">+</mo>
      <mrow xref="m1.6.cmml" id="m1.6d">
        <mi xref="m1.1.cmml" id="m1.1">F</mi>
        <mo xref="m1.6.cmml" id="m1.6e">&ApplyFunction;</mo>
        <mrow xref="m1.6.cmml" id="m1.6c">
          <mo xref="m1.6.cmml" id="m1.6" stretchy="false">(</mo>
          <mi xref="m1.2.cmml" id="m1.2">a</mi>
          <mo xref="m1.6.cmml" id="m1.6a">,</mo>
          <mi xref="m1.3.cmml" id="m1.3">b</mi>
          <mo xref="m1.6.cmml" id="m1.6b" stretchy="false">)</mo>
        </mrow>
      </mrow>
    </mrow>
    <annotation-xml id="m1b" encoding="MathML-Content">
      <apply xref="m1.7" id="m1.7.cmml">
        <plus xref="m1.5" id="m1.5.cmml"/>
        <ci xref="m1.4" id="m1.4.cmml">a</ci>
        <apply xref="m1.6d" id="m1.6.cmml">
          <ci xref="m1.1" id="m1.1.cmml">F</ci>
          <ci xref="m1.2" id="m1.2.cmml">a</ci>
          <ci xref="m1.3" id="m1.3.cmml">b</ci>
        </apply>
      </apply>
    </annotation-xml>
    <annotation id="m1c" encoding="application/x-tex">a+F(a,b)</annotation>
  </semantics>
</math>
\end{lstlisting}
However, this simple example already hints at a hidden complexity. Converting to either
pMML and cMML is straightforward
(given rules for mapping \xmath\ elements to \MathML):
simply walk the tree,
following each \xml{XMRef} to the referenced node and 
choosing the first or second branch of \xml{XMDual} for content or presentation, respectively.
Even cross-linking is straightforward in the absence of \xml{XMDual},
when the generated content or presentation nodes are `sourced' from the same \xmath\ node
($F$, $a$, and $b$, in the example): we simply assign \attr{ID}'s to the source \xmath\ node
and the generated nodes and record the association between the two;
afterwards, the presentation and content nodes that were sourced from the
same \attr{ID} get an \attr{xref} attribute referring to the other, in order to connect them.
But with \xml{XMDual} one has not only to determine when the generated nodes are
related, one has to contend with extra tokens; in the example, the parentheses and comma
appear only in the presentation. Presumably, those tokens should be associated with the
\emph{application} of $F$, as would the containing \xml{m:mrow}?
The desired result is shown in Listing \ref{lst:basic.mathml}.

\begin{lstlisting}[style=xml,label=lst:complex.xml,
  float=p,caption={Internal representation of
    $\left\langle\Psi\middle|\mathcal{H}\middle|\Phi\right\rangle 
     + \defint{a}{b}{F(x)}{x}$}]
<XMApp>
  <XMTok meaning="plus" role="ADDOP">+</XMTok>
  <XMDual>
    <XMApp>
      <XMTok meaning="quantum-operator-product"/>
      <XMRef idref="m2.5"/>
      <XMRef idref="m2.6"/>
      <XMRef idref="m2.7"/>
    </XMApp>
    <XMWrap>
      <XMTok role="OPEN">@$\langle$@</XMTok>
      <XMTok role="ID" xml:id="m2.5">@$\Psi$@</XMTok>
      <XMTok role="CLOSE" stretchy="true">|</XMTok>
      <XMTok role="ID" xml:id="m2.6" font="caligraphic">H</XMTok>
      <XMTok role="OPEN" stretchy="true">|</XMTok>
      <XMTok role="ID" xml:id="m2.7">@$\Phi$@</XMTok>
      <XMTok role="CLOSE">@$\rangle$@</XMTok>
    </XMWrap>
  </XMDual>
  <XMDual>
    <XMApp>
      <XMTok meaning="hack-definite-integral" role="UNKNOWN"/>
      <XMRef idref="m2.1"/>
      <XMRef idref="m2.2"/>
      <XMRef idref="m2.3"/>
      <XMRef idref="m2.4"/>
    </XMApp>
    <XMApp>
      <XMApp>
        <XMTok role="SUPERSCRIPTOP" scriptpos="post2"/>
        <XMApp>
          <XMTok role="SUBSCRIPTOP" scriptpos="post2"/>
          <XMTok mathstyle="display" meaning="integral" role="INTOP">@$\int$@</XMTok>
          <XMTok role="ID" xml:id="m2.1" font="italic">a</XMTok>
        </XMApp>
        <XMTok role="ID" xml:id="m2.2" font="italic">b</XMTok>
      </XMApp>
      <XMApp>
        <XMTok meaning="times" role="MULOP"></XMTok>
        <XMDual xml:id="m2.3">
          <XMApp>
            <XMRef idref="m2.3.1"/>
            <XMRef idref="m2.3.2"/>
          </XMApp>
          <XMApp>
            <XMTok role="FUNCTION" xml:id="m2.3.1" font="italic">F</XMTok>
            <XMWrap>
              <XMTok role="OPEN" stretchy="false">(</XMTok>
              <XMTok role="UNKNOWN" xml:id="m2.3.2" font="italic">x</XMTok>
              <XMTok role="CLOSE" stretchy="false">)</XMTok>
            </XMWrap>
          </XMApp>
        </XMDual>
        <XMApp>
          <XMTok meaning="differential-d" role="DIFFOP" font="italic">d</XMTok>
          <XMTok role="UNKNOWN" xml:id="m2.4" font="italic">x</XMTok>
        </XMApp>
      </XMApp>
    </XMApp>
  </XMDual>
</XMApp>
\end{lstlisting}
A fuller illustration of the issues encountered in typical \LaTeX\ markup
combines complex transfix notations and semantic macros, such as:
\begin{verbatim}
 \left\langle\Psi\middle|\mathcal{H}\middle|\Phi\right\rangle 
  +  \defint{a}{b}{F(x)}{x} 
\end{verbatim}
This example, whose internal form is shown in Listing \ref{lst:complex.xml},
involves quantum-mechanics notations,
which \LaTeXML's parser is happily able to recognize. Additionally,  we've introduced a semantic
macro \verb|\defint| to represent definite integration, which will be transformed to
so-called `Pragmatic' Content MathML form, to enhance the illustration with a many-to-many
correspondence.  (The implementation of \verb|\defint| is not difficult, but outside
the scope of this article)

\section{Algorithm}\label{algorithm}
The goal is to associate each of the generated target pMML (or cMML) nodes
with some node(s) in the generated cMML (or pMML, respectively).
We do this by ascribing to each generated node a \emph{source} \xmath\ node,
not necessarily the \emph{current} node, the one that directly generated the target node.
Once this is done, the cross-referencing is easily established: the \attr{xref} of
a pMML (cMML) node is the cMML (pMML, respectively) node that was ascribed to the
same source \xmath\ node; if multiple nodes were ascribed to that source node,
the first target node, in document order, is the sensible choice.

A key to deciding which \xmath\ node to ascribe as the source is whether
the node is visible to either or both branches. The common, simple, case is
an \xmath\ node, visible to both branches, that generates a token node in the target;
in that case the current node is used as the source.
Node visibility can be determined by an algorithm such as the marking part of mark-and-sweep garbage
collection.

However, \MathML\ elements which are containers generally do not correspond to symbols,
and ought to be associated with the nearest application 
(think \xml{m:apply} or \xml{m:mrow})\footnote{
Exceptions are \xml{m:msqrt} or \xml{m:menclose} where they tend to represent
\emph{both} the application of an operation and yet are the only visible manifestation of 
the operator! However, we also note that a common use of cross-linking in \HTML\ is to
turn them into href links; but \HTML\ does not allow nested links!}.
In this case, the source should be the nearest ancestor \xml{XMDual}
of the current \xmath\ node, which we'll call the \emph{current container}.

Similar reasoning applies in the special case when a token symbol (non-container) is
generated from an \xmath\ token which is not visible to the opposite branch;
it may simply be notational icing of some transfix, or it may be the only visible manifestation
of what we'll call the \emph{current operator}.  The current operator is the top-most operator
being applied within the current container.
In the example, the angle brackets and vertical bars are the only visible manifestation
of the \texttt{quantum-operator-product} operator.

In summary, the \emph{source} node for a given \emph{target} is
\begin{lstlisting}[escapechar=@]
if @\emph{target}@ is a container
  if @\emph{current container}@ exists
    @\emph{current container}@
  else
    @\emph{current}@   
else @\emph{target}@ is visible in both branches
  @\emph{current}@
else if @\emph{current container}@ exists
  if @\emph{current operator}@ is hidden from presentation
    @\emph{current operator}@
  else
    @\emph{current container}@
else
  @\emph{current}@
\end{lstlisting}
Applying this method to the example from Listing \ref{lst:complex.xml} yields
\ref{lst:complex.mathml}, where we can see, for example, that the
angle brackets and vertical bars are associated with the \texttt{quantum-operator-product}
operator while the various \xml{m:bvar}, \xml{m:lowlimit}, etc, are properly associated
with the integral, \emph{not} the integral operator.
\begin{lstlisting}[style=xml,label=lst:complex.mathml,
  float=p,caption={\MathML\ representation of
    $\left\langle\Psi\middle|\mathcal{H}\middle|\Phi\right\rangle 
     + \defint{a}{b}{F(x)}{x}$}]
<math display="block" alttext="..." class="ltx_Math" id="m2">
  <semantics id="m2a">
    <mrow xref="m2.13.cmml" id="m2.13">
      <mrow xref="m2.9.cmml" id="m2.9">
        <mo xref="m2.8.cmml" id="m2.8">&LeftAngleBracket;</mo>
        <mi mathvariant="normal" xref="m2.5.cmml" id="m2.5">&Psi;</mi>
        <mo xref="m2.8.cmml" id="m2.8a" stretchy="true" fence="true">|</mo>
        <mi xref="m2.6.cmml" id="m2.6" class="ltx_font_mathcaligraphic">&HilbertSpace;</mi>
        <mo xref="m2.8.cmml" id="m2.8b" stretchy="true" fence="true">|</mo>
        <mi mathvariant="normal" xref="m2.7.cmml" id="m2.7">&Phi;</mi>
        <mo xref="m2.8.cmml" id="m2.8c">&RightAngleBracket;</mo>
      </mrow>
      <mo xref="m2.10.cmml" id="m2.10">+</mo>
      <mrow xref="m2.12.cmml" id="m2.12c">
        <msubsup xref="m2.12.cmml" id="m2.12">
          <mo xref="m2.11.cmml" id="m2.11" symmetric="true" largeop="true">&int;</mo>
          <mi xref="m2.1.cmml" id="m2.1">a</mi>
          <mi xref="m2.2.cmml" id="m2.2">b</mi>
        </msubsup>
        <mrow xref="m2.12.cmml" id="m2.12b">
          <mrow xref="m2.3.cmml" id="m2.3c">
            <mi xref="m2.3.1.cmml" id="m2.3.1">F</mi>
            <mo xref="m2.3.cmml" id="m2.3d">&ApplyFunction;</mo>
            <mrow xref="m2.3.cmml" id="m2.3b">
              <mo xref="m2.3.cmml" id="m2.3" stretchy="false">(</mo>
              <mi xref="m2.3.2.cmml" id="m2.3.2">x</mi>
              <mo xref="m2.3.cmml" id="m2.3a" stretchy="false">)</mo>
            </mrow>
          </mrow>
          <mo xref="m2.11.cmml" id="m2.11a">&InvisibleTimes;</mo>
          <mrow xref="m2.12.cmml" id="m2.12a">
            <mo xref="m2.11.cmml" id="m2.11b">d</mo>
            <mi xref="m2.4.cmml" id="m2.4">x</mi>
          </mrow>
        </mrow>
      </mrow>
    </mrow>
    <annotation-xml id="m2b" encoding="MathML-Content">
      <apply xref="m2.13" id="m2.13.cmml">
        <plus xref="m2.10" id="m2.10.cmml"/>
        <apply xref="m2.9" id="m2.9.cmml">
          <csymbol xref="m2.8" id="m2.8.cmml" cd="latexml">quantum-operator-product</csymbol>
          <ci xref="m2.5" id="m2.5.cmml">normal-&Psi;</ci>
          <ci xref="m2.6" id="m2.6.cmml">&HilbertSpace;</ci>
          <ci xref="m2.7" id="m2.7.cmml">normal-&Phi;</ci>
        </apply>
        <apply xref="m2.12c" id="m2.12.cmml">
          <int xref="m2.11" id="m2.11.cmml"/>
          <bvar xref="m2.12c" id="m2.12a.cmml">
            <ci xref="m2.4" id="m2.4.cmml">x</ci>
          </bvar>
          <lowlimit xref="m2.12c" id="m2.12b.cmml">
            <ci xref="m2.1" id="m2.1.cmml">a</ci>
          </lowlimit>
          <lowupper xref="m2.12c" id="m2.12c.cmml">
            <ci xref="m2.2" id="m2.2.cmml">b</ci>
          </lowupper>
          <apply xref="m2.3c" id="m2.3.cmml">
            <ci xref="m2.3.1" id="m2.3.1.cmml">F</ci>
            <ci xref="m2.3.2" id="m2.3.2.cmml">x</ci>
          </apply>
        </apply>
      </apply>
    </annotation-xml>
    <annotation id="m2c" encoding="application/x-tex">...</annotation>
  </semantics>
</math>
\end{lstlisting}

\section{Outlook}
% An interesting contrast, somewhat contrived, is integrals
% where the `excess' symbols are on the content side, at least in `pragmatic' cmml.
% Note that we currently have not implemented sufficient semantic analysis
% to completely recognize integrals marked up in traditional presentation style,
% i.e. extracting the differentials, recognizing the nature of the integration
% limits or path. An interesting challenge will be to construct the parallel
% markup for the integrand without the differential for the content branch,
% while correlating the individual symbols and structures with the
% presentation branch, with the differential embedded in its original layout
% (Think of a common expression such as $\int \frac{dx}{x}$).

The Digital Library of Mathematical Functions\footnote{\url{http://dlmf.nist.gov/}}
had from the outset linkage from (most) symbols to their definitions. However this new approach
to the problem provides a much cleaner implementation, and allowed the mechanism
to be extended to less textual operators such as binomials, floor, 3-j symbols, etc.

Parallel markup must also be adapted to larger structures such as eqnarray, 
and AMS alignments with intertext containing multiple formula
and/or document-level text markup.  While the fundamental issue is the
same --- separating presentation and content forms --- this seems to demand a
distributed markup that separates the presentation and content forms into
distinct math containers.   \LaTeXML\ currently has an ad-hoc, but not entirely
satisfactory solution for this, but we will experiment with adapting the methods
described here.  However, it remains to be seen whether cross-referencing across
separate math containers can be made useful.

And, now that generating Content MathML is more fun, we must continue work
towards generating \emph{good} Content MathML. Ongoing work will attempt to establish
appropriate OpenMath Content Dictionaries, probably in a FlexiFormal sense\cite{Kohlhase:tffm13},
improved math grammar, and exploring semantic analysis.

\bibliography{paper}
\end{document}